\documentclass[letterpaper]{IAU-TrA}
\usepackage{graphicx}

\title[CLOSE BINARY STARS]     
{}

\author[COMMISSION 42]   
{}

\pubyear{2009}
\volume{Volume XXVIIA}
\pagerange{001--008}

\jname{Transactions IAU, Volume XXVIIA}
\editors{Karel A. van der Hucht, ed.}
\begin{document}

\maketitle

{\bf

\large
\begin{tabbing}
\hspace*{64mm}       \=                                               \kill
COMMISSION~42        \> CLOSE BINARY STARS                            \\[0.5ex]
                     \> {\small\it \a'{E}TOILES DOUBLES SERR\a'{E}ES} \\
                     \>                                               \\
\end{tabbing}

\normalsize

\begin{tabbing}
\hspace*{64mm}       \=                                               \kill
PRESIDENT            \> Slavek M. Rucinski                            \\
VICE-PRESIDENT       \> Ignasi Ribas                                  \\
PAST PRESIDENT       \> Alvaro Gim\a'{e}nez                           \\ 
ORGANIZING COMMITTEE \> Petr Harmanec, Ronald W. Hilditch,            \\ 
                     \> Janusz Kaluzny, Panayiotis Niarchos,          \\
                     \> Birgitta Nordstr\"om, Katalin Ol\a'{a}h,      \\
                     \> Mercedes T. Richards, Colin D. Scarfe,        \\   
                     \> Edward M. Sion, Guillermo Torres,             \\
                     \> Sonja Vrielmann                               \\
\end{tabbing}

\bigskip

\noindent
TRIENNIAL REPORT 2006-2009
}

\firstsection 

\section{Introduction}

Two meetings of interest to close binaries took place during the reporting 
period: A full day session on short-period binary stars -- mostly CV's -- 
(\cite[Milone et al. 2008]{Milone2008}) during the 2006 AAS Spring meeting in 
Calgary and the very broadly designed IAU Symposium 240 in Prague in 2006, 
with many papers on close binaries \cite[(Hartkopf et al.\ 
2007)]{Hartkopf:07}. In addition, the book by \cite{Eggleton2006}, which is 
a comprehensive summary of evolutionary processes in binary and multiple 
stars, was published.

The report that follows consists of individual contributions of the 
Commission 42 Organizing Committee members. Its goal has been to give very 
personal views of a few individuals who are active in the field, so the 
report does not aim at covering the whole field of close binaries.

\section{Close binary research from the perspective of BCB (C.~D.\ Scarfe)}

Some light can be thrown on the state of research on close binaries by 
considering the papers listed in the semi-annual Bibliography of Close 
Binaries (BCB). Since no similar summary appeared in the last few 
triennial reports of the Commission, we review here the patterns seen in 
the last 14 issues, since the most recent change of editorship in 2001, 
but with emphasis on the most recent six issues.

The total output of work on close binaries, appears to have decreased 
slightly during this decade. The two largest recent issues were those of 
December 2001 and June 2003, and the two smallest were those of December 
2005 and December 2006. Part of this decline can be attributed to the 
editors' efforts to list only once, in the ``Collections of Data'' 
section, papers giving the same kind of results for numerous binary 
systems. This avoids repetition of the listing for each system in the 
``Individual Stars'' section.

There is a slow trend toward larger numbers of authors per paper. The mean 
number of authors per paper has grown from 3.80 for the 420 entries in 
\#72 to 4.88 for the 380 entries in \#85. These means are affected by the 
few papers with very large numbers of authors -- over 100 in rare cases - 
but the medians show the same trend. There is also a consistent systematic 
difference between ``Individual Star'' (mostly observational) papers and 
those in the ``General'' (mainly theoretical) category, the latter having 
about two fewer authors than the former, on average.

The literature of close binary stars, like that of other variables, 
suffers from a multiplicity of names for the objects of study. We have 
tried to follow the normal variable-star nomenclature of letters or 
numbers followed by constellation names, or their numbers in widely used 
catalogues such as the HD catalogue. Failing that, we prefer a 
coordinate-based nomenclature, and do our best to avoid other names. We 
encourage authors of papers to use standard nomenclature, making use of 
the SIMBAD identifier lists.
 
We turn now to the individual objects that are of abiding interest, or 
have attracted great temporary interest, based on the number of papers 
about them in the most recent 14 issues of BCB. Consistent with the 
preceding paragraph, we refer to each object by the name preferred for 
BCB, usually the standard variable-star designation. Papers have been 
listed in every issue for only two objects: V1357~Cyg (77 papers) and 
V1487~Aql (62 papers). Three other objects have more than 30 papers 
listed: KV~UMa with 37 in 12 issues, V1343~Aql with 34 in 11, and HZ~Her, 
with 33 in 12. However, PSR J0737--3039 was vigorously studied between 
2004 and 2006, resulting in 29 papers in only 6 issues, including a record 
number of 12 in a single issue (\#80). The remaining objects with at least 
20 papers are V381~Nor (25 in 11), V615~Cas (24 in 9), V821~Ara (24 in 
12), WZ~Sge (23 in 10), V1033~Sco (23 in 9), V1521~Cyg and BR~Cir (both 22 
in 12), V818~Sco (21 in 10) and V4580~Sgr (20 in 11). Clearly, X-ray 
binaries dominate this list, but objects as the recurrent nova RS~Oph, 
CH~Cyg, AE~Aqr and $\eta$~Car are close behind (averaging $>1$ paper per 
issue). All but RS~Oph are of continuing interest, whereas that object's 
recent outburst has led to more papers in the latest three issues than for 
any other object.

\section{Low-mass binaries and model discrepancies (G.\ Torres)}

The properties of short-period double-lined eclipsing binaries (EBs) with 
M-type components have long been known to present disagreements compared 
to predictions from stellar evolution models (e.g., \cite[Popper 
1997]{Popper:97}), particularly in their radii and effective temperatures.  
Since mid-2006 there have been considerable efforts to find additional 
low-mass systems in order to test theory, such as the Monitor Project that 
focuses on young open clusters (\cite[Aigrain et al. 2007]{Aigrain:07}).  
A number of other discoveries of potentially useful systems have occurred 
serendipitously or as a result of searching existing photometric databases 
(e.g., \cite[Bayless \& Orosz 2006]{Bayless:06}; \cite[Young et al. 
2006]{Young:06}; \cite[L\'opez-Morales \& Shaw 2007]{Lopez-Morales:07a}; 
\cite[Vaccaro et al. 2007]{Vaccaro:07}; \cite[Becker et al. 
2008]{Becker:08}). Most of these systems continue to indicate that the 
sizes of M dwarfs are larger than predicted by theory, and their 
temperatures are lower. A few single-lined F+M systems resulting from 
searches for transiting planets have also provided useful comparisons 
(e.g., \cite[Beatty et al. 2007]{Beatty:07}). The discrepancies with 
models are not confined to low-mass systems, however, and have been shown 
to extend up to masses almost as large as the Sun (\cite[Torres et al. 
2006]{Torres:06}).

Chromospheric activity that is common in late-type and tidally locked 
binaries has long been suspected to be the culprit, as pointed out by many 
of the above investigators, and observational evidence for this has been 
mounting recently (see \cite[L{\'o}pez-Morales 2007]{Lopez-Morales:07b}; 
\cite[Reiners et al. 2007]{Reiners:07}, \cite[Morales et al. 
2008]{Morales:08}). Both heavy spot coverage and reduced convective 
efficiency due to strong magnetic fields may play a role. The problem has 
been reviewed, e.g., by \cite{Ribas:07}. Although much remains to be done 
both observationally and on the theoretical side, recent modelling studies 
of the evolution of low-mass stars under the influence of chromospheric 
activity are very encouraging. They appear to be able to explain the 
observed effects quite well with reasonable magnetic field strengths and 
typical spot filling factors (\cite[Chabrier et al. 2007]{Chabrier:07}). 
Similar progress has been made in our understanding of the origin and 
strength of magnetic fields in fully convective stars that lack a 
tachocline (\cite[Browning 2008]{Browning:08}).

\section{W~UMa-type binaries (S.~M.\ Rucinski)}

Although numbers of publications remain high, the field of contact 
binaries does not enjoy any obvious progress. The literature is still 
abundant in single-object photometric solutions of dubious usefulness (is 
anybody going ever to combine them?), with unrealistically small parameter 
uncertainties; they contribute little and just inflate citation lists. 
While very few new theoretical studies have been done -- with some notable 
exceptions, e.g., the model of AW~UMa (\cite[Paczy\'nski et al. 
2007]{Paczynski:07}), a model of evolution including the mass reversal 
(\cite[St\c epie\'n 2006a]{Stepien:06a}) -- most efforts concentrate on 
observations and their interpretation.

New, shallow but extensive surveys of variability such as ASAS 
(\cite[Paczy\'nski et al. 2006]{BP2006}) have resulted in large numbers of 
contact binaries to about $V \simeq 12$ mag. W~UMa type binaries are not 
very common, with one per about 500 late-type dwarfs ($\simeq 1.0 \times 
10^{-5}$ pc$^{-3}$; \cite[Rucinski 2006]{Rucinski:06}), but they are easy 
to discover, hence their strong representation in variable star 
catalogues. The catalogues are biased in terms of physical properties: an 
elementary correction for the sample volume shifts the peak of the period 
distribution from 0.35~d to 0.27~d (\cite[Rucinski 2007]{Rucinski:07}). 
This is not far from the sharp cut-off at 0.22~d, where CC~Comae is 
accompanied by the new shortest-period record holder discovered by
ASAS, GSC~01387--00475 (\cite[Rucinski \& Pribulla 2008]{Rucinski:08}). 
\cite{Stepien:06b} explains the 
cut-off by the strong dependence of the angular-momentum-loss efficiency 
on mass, $-dH/dt \propto M^3$, so that low-mass objects simply have had no 
time to evolve beyond it.

We know now that W~UMa binaries almost certainly have companions 
(\cite[Rucinski et al. 2007]{RPvK:07}; and citations therein), in full 
confirmation of the work of \cite{tokov2006} of the increasing 
multiplicity at short binary periods.

Only AW~UMa is worth mentioning as an individual object: This flagship 
contact binary, when analyzed spectroscopically in detail, appears not to 
be in contact at all, but is a complex and hard to interpret semi-detached 
binary (\cite[Pribulla \& Rucinski 2008]{PR2008}). This teaches us a 
lesson that light curve fits even with most elaborate synthesis codes may 
yield an entirely incorrect picture. There is also a major question if the 
contact model applies to W~UMa binaries at all...

Spectroscopy of bright contact binaries is essential for progress. Faint, 
accidentally discovered systems continue to be observed photometrically 
because this is easy, but usefulness of this is questionable. Note that 
hundreds W~UMa's still remain to be detected to $V=10$ mag, but most of 
them have small amplitudes because of low inclinations and/or dilution by 
the light of companions.

\section{Cataclysmic Variables (E.~M.\ Sion)}

Since the discovery of the first CV containing a pulsating white dwarf 
with ZZ~Ceti-like non-radial $g$-mode oscillations, the list of such 
cataclysmic white dwarf pulsators has grown to 12 (\cite[Gaensicke et al. 
2006]{Gaensicke2006}; \cite[Mukadam et al. 2007]{Mukadam2007}). These 
objects are being probed with asteroseismological techniques to address 
the extent to which accretion affects the white dwarf mass, temperature, 
and composition and how efficiently angular momentum is transferred into 
the core.

Classical novae and dwarf novae are powered by two entirely different 
outburst mechanisms and have never been seen closely spaced in time. 
However, \cite{Sokoloski2006} discovered the first example (Z~And) of an 
accretion disk instability depositing enough material that a 2 magnitude 
flare was triggered, followed by a thermonuclear explosion on an accreting 
white dwarf which brightened the system to $10^4\,L_\odot$, accompanied by 
a mass ejection. The two closely spaced events occurred in a symbiotic 
variable, but this so-called ``combination nova'' holds potential 
implications for CVs as well. A likely classical nova shell has been 
detected around the prototype dwarf nova Z~Cam by \cite{Shara2007}.  This 
would be one of the first direct links between classical novae and dwarf 
novae.

\cite{Darnley2006} estimate that if recent X-ray surveys of the galactic 
plane are correct in predicting more than $>2 \times 10^4$ CVs with X-ray 
luminosity $<2 \times 10^{33}$ erg/s, then each must undergo a nova 
explosion once per millennium to keep up the expected nova rate of 34 per 
year. \cite{Shara2006} reported a new class of CVs, found exclusively in 
globular clusters, whose members never went through a common envelope 
binary phase. They predict that their distribution of orbital periods 
should not have the standard CV period gap at 2--3 hours. 
\cite{Pretorius2008} presented an independent new sample of CVs, selected 
by $H\alpha$ emission. They cannot reconcile the large ratio of short- to 
long-period CVs predicted by standard CV evolution theory with their 
sample unless the rate of angular momentum loss below the period gap is 
increased by a factor of at least 3. \cite{Schmidtobreick2006} reported 
that there are now 11 known CVs in the period gap, but a shortage of 
expected dwarf novae that have evolved past the turn-around in orbital 
period. This again suggests that a mechanism of angular momentum loss in 
addition to gravitational radiation is required. Recent studies of the 
temperature distribution of CV white dwarfs versus orbital period below 
the period gap also suggest additional angular momentum loss.

\cite{Knigge2007} has introduced a valuable method for determining the 
distances to cataclysmic variables using 2MASS $JHK$ photometry and a 
semi-empirical relationship between donor absolute magnitude in the 
$K$-band and orbital period. The long sought-after and controversial mass 
of the white dwarf in WZ~Sge was determined by \cite{Steeghs2007} in 2004. 
The WD mass is $0.85\pm0.03$~M$_\odot$. Finally, \cite{Kromer2007} have 
succeeded in constructing model accretion disks with emission lines formed 
through irradiation of the inner disk by the hot white dwarf. This is a 
significant breakthrough since these accretion disk models enable emission 
line modelling {\it ab initio\/}.

\section{Algol-type binaries (M.~T.\ Richards)}

The important fundamental light curve and spectroscopic analyses of 
several systems has continued (e.g., \cite[Angione \& Sievers 2006]{AS06}; 
\cite[Soydugan et al.\ 2007]{Sea07}) We also need additional studies of 
infrared light curves like those of \cite{Lea06}.  Analyses that include a 
third body (e.g., \cite[Li 2006]{L06}; \cite[Hoffman et al.\ 2006]{Hea06}) 
suggest that many Algols have a companion in a long-period orbit and that 
the third body has a non-negligible effect on the binary through long-term 
variations in the binary orbital period.  Both light curves and spectra 
have been combined by \cite{vHW07} to refine the derivation of third-body 
parameters and to challenge the existence of a third body in VV~Ori. The 
six largest coronal X-ray flares ever detected by {\it Chandra} were 
studied by \cite{NB07} to derive the flare properties.  Flare size and 
position have been constrained by studying eclipsed X-ray flares on Algol 
and VW~Cep (\cite[Sanz-Forcada et al.\ 2007]{SFea07}).

The angular momentum evolution of 74 detached binaries and 61 semidetached 
binaries was studied by \cite{Iea06}. Evolutionary models suggest that a 
circumbinary disk could extract angular momentum from the binary, thereby 
causing the orbit to shrink. This disk may explain the low mass ratio 
systems undergoing rapid mass transfer (\cite[Chen et al.\ 2006]{Cea06}).  
Conservative Roche Lobe overflow explains the observed range of orbital 
periods in systems with B-type primaries but not for those with high mass 
ratios (\cite[van Rensbergen et al.\ 2006]{vRea06}). However, hot spots and 
the spin-up of the primary may lead to significant mass loss.

Synthetic spectra of cool Algol secondaries can now be calculated with the 
{\sc LinBrod} code (\cite[Bitner \& Robinson 2006]{BR06}).  In addition, 
synthetic spectra of the accretion disk and gas stream were calculated for 
TT~Hya using the {\sc shellspec} code (\cite[Miller et al. 2007]{Mea07}).  
The physical properties of the accretion disk and stream were derived from 
the direct comparison between the observed and synthetic spectra of the 
system and also by using Doppler tomography to demonstrate visually that 
the accretion structures have been properly modelled.

$K$-band direct imaging of SS~Lep with VINCI/VLTI and photometry from the 
UV to far-IR has been used by \cite{Vea07} to reveal the stars and a dust 
shell or disk.  Indirect images of W~Cru based on the eclipse mapping 
method identified a clumpy disk structure (\cite[Pavlovski et al.\ 
2006]{Pea06}).  Finally, new 3D Doppler tomograms of the U~CrB binary 
demonstrate that the alternating accretion disk and stream structures have 
significant flow velocities beyond the central 2D velocity plane and there 
is evidence of a gas jet at the star-stream impact site 
\cite[Agafonov et al.\ (2006)]{Agafonov:06}.

\section{The oEA stars (P.~G.\ Niarchos)}

Following the introduction of the name ``oEA'' (oscillating EA) for (B)A-F 
spectral type, mass-accreting, main-sequence, pulsating stars in 
semi-detached Algol-type EBs by \cite{Mkrtichian2004}, several studies 
having been published in the recent years.  The oEA stars are the former 
secondaries of evolved, semi-detached EBs which are (still) undergoing 
mass transfer and form a class of pulsators close to the main-sequence.

\cite{Soydugan2006a}, considering a sample of 20 EBs with $\delta$~Sct 
primaries, discovered that there is a possible relation among the 
pulsational periods of the primaries and the orbital periods of the 
systems. An important contribution was made by \cite{Soydugan2006b}, who 
presented a catalogue of close binaries (25 known and 197 candidate 
binaries with pulsating components) located in the $\delta$~Sct region of 
the instability strip. A status report on the search for pulsations in 
primary components of Algol-type systems was presented by 
\cite{Mkrtichian2006}, while a revised list of presently known oEA stars 
and a discussion on the pulsation mode visibility and strategies of mode 
identification was published by \cite{Mkrtichian2007}.

Studies of individual systems: The following oEA systems were studied: 
AB~Per, QU~Sge, RZ~Cas, IV~Cas, Y~Cam, and CT~Her. Moreover, 
\cite{Pigulski2007} and \cite{Michalska2007} detected pulsating components 
in 25 EBs in the ASAS-3 database.

\section{Effects of binarity on stellar activity (K.\ Ol\'ah)}

\cite{K07} present a detailed spot modeling analysis and Doppler images 
for the primary stars of the RS CVn-type binary $\zeta$~And. The 
photometric light modulation originates in the distorted geometry of the 
primary, and additionally, comes from spots, which preferably appear on 
the stellar surface towards the companion star and opposite to it. Doppler 
maps also show low latitude spots with a temperature contrast of about 
1000~K, and some weak polar features. Weak solar-type differential 
rotation was derived from the cross-correlation of the consecutive Doppler 
maps.

\cite{M08} reported observational evidence of interacting coronae of the 
two components of the young binary system V773~Tau~A. The VLBA and 
Effelsberg radio images show two distinctive structures, which are 
associated with each star. In one image the two features are extended up 
to $18$~R$_\star$ each and are nearly parallel revealing the presence of 
two interacting helmet streamers, observed for the first time in stars 
other than the Sun. During the stellar rotation, these helmet streamers 
come into collision producing periodical flares. This is the first 
evidence that even if the flare origin is magnetic reconnection due to 
inter-binary collision, both stars independently emit in the radio range 
with structures of their own. The helmet streamers appear to interact 
throughout the whole orbit, although the radio flares become stronger when 
the stars approach. Around periastron the stellar separation is only 
$30$~R$_\star$, where the two streamers overlap producing the observed 
giant flares.

\cite[Dunstone et al.\ (2008)]{D08} present the first measurements of 
surface differential rotation of the young, pre-main sequence binary 
system HD~155555. Both components are found to have high rates of 
differential rotation, similar to those of the same spectral type Main 
Sequence single stars. The results for HD~155555 are therefore in contrast 
to those found in other, more evolved binary systems where negligible or 
weak differential rotation has been discovered. The rotation of both stars 
of HD~155555 is synchronous and the system is tidally locked. The authors 
found that more likely the convection zone depth is the cause of the low 
differential rotation rates of evolved giants, rather than the effects of 
tidal forces. The strong differential rotation provides extra stresses on 
the fields, and the reconnection of these long binary field loops 
significantly contribute to the X-ray luminosity of the system and to the 
elevated frequency of large flares.

\vspace{3mm}

{\hfill Slavek Rucinski and Ignasi Ribas}

{\hfill {\it President and V-President of Commission 42}}

\end{document}